\documentclass[12pt]{elsarticle}
\usepackage{graphicx} 
\usepackage{float}
\usepackage{subcaption}
\usepackage{tikz}
\usepackage{hyperref}
\usepackage{multirow}
\usepackage{booktabs}
\usetikzlibrary{positioning, arrows.meta}
\usepackage[a4paper,margin=2.5cm]{geometry}

\begin{document}
\begin{frontmatter}

\title{Structural Analysis of Journal Columns Using Ordinal Patterns and Information-Theoretic Measures}
\author{Cynthia Z. Zhou-Lin$^{1}$, Jos\'e L. Herrera-Diestra$^{2}$, Luciano Stucchi$^{1,*}$\\[0.5em]
$^{1}$ Universidad del Pacífico, Lima, Per\'u\\
$^{2}$ Tarleton State University, Texas, United States\\
$^*$stucchi\_l@up.edu.pe
}


\begin{abstract}
Source attribution in journalistic text is typically approached through semantic representations, such as term weighting or neural embeddings. These approaches capture topical content, but usually overlook how a text is organized as a sequence. Here, we show that purely structural, content-independent features can discriminate between news sources on their own. We converted articles from fourteen Peruvian online newspapers into numerical sequences using two independent encodings, word length and lexical frequency. Then, we analyzed each of them through ordinal pattern analysis, computing pattern and transition probabilities, permutation entropy, disequilibrium, and statistical complexity. Both encodings yield non-uniform ordinal pattern distributions, well-defined preferential transitions, and coherent entropy complexity signatures across sources. Unsupervised clustering recovers three coherent groups in the feature space, and supervised classifiers trained on these features achieve accuracy up to 0.99, which remains stable even under substantial feature reduction. The consistency of these results across two structurally unrelated encodings strongly indicates that ordinal features capture a genuine dynamical fingerprint of each source's writing style rather than an artifact of either representation, which extends this framework to Spanish-language journalism, a setting that remains comparatively underexplored in complexity-based text analysis.
\end{abstract}

\end{frontmatter}

\textbf{Keywords:} Ordinal patterns; Permutation entropy; Statistical complexity; Stylometry; Source attribution; Journalistic text

\section{Introduction}

The rapid expansion of online news media has increased the demand for automated methods to analyze journalistic text across large collections. A central task in this area is source attribution, understood as identifying the newspaper or outlet responsible for a given article, with applications in media monitoring, bias detection, and verification of information provenance \cite{Sebastiani2002, Manning1999}. Existing approaches typically rely on semantic representations, such as bag-of-words models and TF-IDF weighting \cite{salton1988, manning2008}, or more recently on neural embeddings \cite{mikolov2013, devlin2019}, which capture topical content but give limited attention to structural and stylistic organization. Yet, each publication tends to develop recognizable writing habits reflected in consistent patterns of vocabulary richness, sentence rhythm, and lexical repetition. These patterns leave structural traces that remain detectable in the text regardless of the article's subject matter.
  
This intuition underlies the field of computational stylometry, which seeks to attribute authorship or source through a quantitative analysis of writing style \cite{Stamatatos2009}. Features such as word length distributions, function word frequencies, and lexical diversity measures have proven effective for authorship attribution across literary and journalistic corpora \cite{Koppel2009}. These structural descriptors are useful because they operate below the semantic level, so that a newspaper's structural fingerprint persists across topics, reporters, and time periods, providing a robust signal for source identification. Natural language exhibits well-known regularities at this structural level, including power-law word frequency distributions \cite{Zipf1949} and characteristic word length profiles shaped by cognitive and communicative constraints \cite{Sigurd2004}, both of which vary measurably across publications and writing styles. Ordinal pattern analysis, introduced by Bandt and Pompe \cite{bandt2002}, offers an efficient framework for characterizing the structure of numerical sequences. The method encodes only the relative ordering of consecutive values rather than their absolute magnitudes, producing a symbolic representation that is robust to noise, insensitive to monotonic transformations, and free of distributional assumptions. When a text is converted to a numerical sequence by replacing each word with its character length or its relative frequency within the document, ordinal patterns extracted from that sequence capture regularities in lexical flow, rhythmic organization, and structural variability that semantic approaches cannot reach.

Several information-theoretic descriptors can be derived from the distribution of ordinal patterns. Permutation entropy \cite{bandt2002} quantifies the degree of randomness in the symbolic sequence. Disequilibrium \cite{LopezRuiz1995} measures the deviation of the pattern distribution from uniformity. Statistical complexity \cite{rosso2007}, defined as the product of entropy and disequilibrium, captures the balance between randomness and organization that characterizes structured complex systems. These measures jointly place each document in the entropy-complexity plane, allowing structural properties to be compared across texts without reference to semantic content \cite{rosso2007}. Ordinal pattern methods have been applied in biomedical signal processing and financial time series analysis \cite{Zanin2012}, as well as in the study of dynamical systems \cite{rosso2007}. Their use in natural language processing is more recent, and their application to source attribution in non-English journalistic corpora has not been systematically explored. To the best of our knowledge, Spanish-language news media have received little attention in complexity-based text analysis, despite the considerable linguistic and stylistic variation across the Spanish-language press, suggesting that such corpora may offer an informative test case for this type of structural analysis.

In this paper, we develop a framework based on ordinal patterns and information-theoretic measures to analyze and classify news articles from Peruvian online newspapers written in Spanish. Each document is converted to a numerical sequence using two encodings, word length and lexical frequency, from which ordinal patterns are extracted with embedding dimension $D = 3$ and delay $\tau = 1$. Pattern probability distributions, transition probability matrices, permutation entropy, disequilibrium, and statistical complexity are then computed to form a feature vector for each article. We evaluate these representations through unsupervised clustering with K-means and PCA visualization, and through supervised classification using Logistic Regression, Random Forest, and a Multilayer Perceptron \cite{pedregosa2011}. The results show that ordinal features encode a distinctive structural fingerprint for each source, with classification accuracy reaching up to 0.99 for the best-performing model and remaining above 0.94 across all evaluated classifiers, confirming that newspaper attribution can be carried out from structural sequence properties alone.

The remainder of this paper is organized as follows. Section II describes the methodology, including corpus collection, preprocessing, numerical encodings, ordinal pattern extraction, and feature construction. Section III presents experimental results for clustering and classification. Section IV discusses the findings and Section V concludes the paper.

\section{Methodology}

The proposed methodology was evaluated on a corpus of 1,261 news articles collected through web scraping from 14 Peruvian online newspapers. Depending on the availability of articles from each source, the corpus contains between 35 and 100 documents per newspaper. For each article, the dataset includes the newspaper source, title and full text, providing a diverse benchmark for evaluating structural representations across different editorial styles. The complete list of newspapers and their approximate access dates are provided in Table~\ref{tab:S1}. The dataset are publicly available at the project repository \cite{github_repo}.

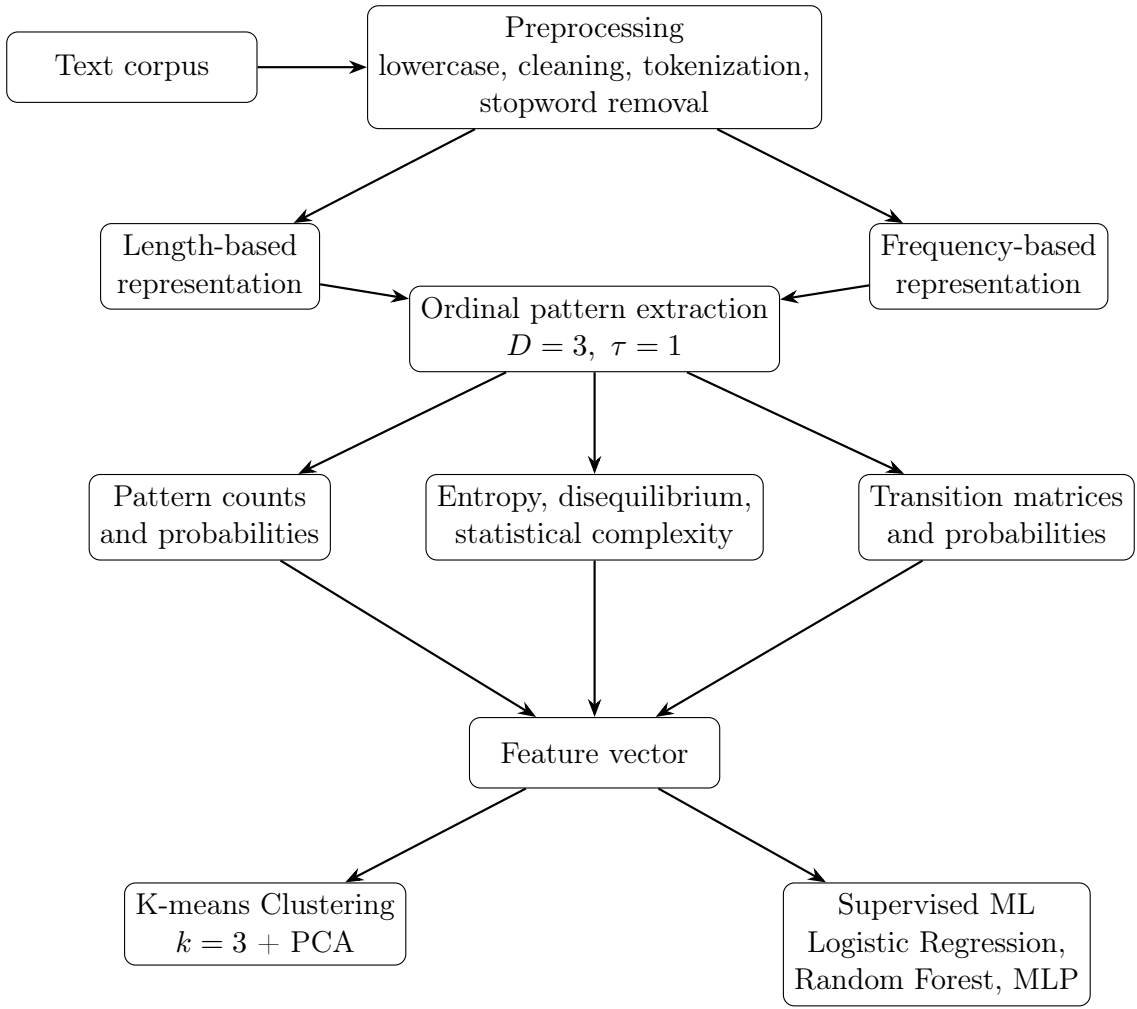
\begin{figure}[H]
\centering
\resizebox{0.95\textwidth}{!}{
\begin{tikzpicture}[
    node distance=1.2cm and 1.4cm,
    every node/.style={font=\small},
    box/.style={
        rectangle,
        rounded corners,
        draw=black,
        align=center,
        minimum width=3.2cm,
        minimum height=0.9cm
    },
    smallbox/.style={
        rectangle,
        rounded corners,
        draw=black,
        align=center,
        minimum width=2.8cm,
        minimum height=0.8cm
    },
    arrow/.style={->, thick, >=Stealth}
]

\node[box] (corpus) {Text corpus};
\node[box, right=of corpus] (prep) {Preprocessing\\
lowercase, cleaning, tokenization,\\ stopword removal};

\node[smallbox, below left=1.2cm and 0.6cm of prep] (length) {Length-based\\representation};
\node[smallbox, below right=1.2cm and 0.6cm of prep] (freq) {Frequency-based\\representation};

\node[box, below=2.0cm of prep] (ordinal) {Ordinal pattern extraction\\
$D=3,\ \tau=1$};

\node[smallbox, below left=1.3cm and 1.0cm of ordinal] (static) {Pattern counts\\
and probabilities};

\node[smallbox, below right=1.3cm and 1.0cm of ordinal] (dynamic) {Transition matrices\\
and probabilities};

\node[smallbox, below=1.3cm of ordinal] (info) {Entropy, disequilibrium,\\
statistical complexity};

\node[box, below=2.0cm of info] (features) {Feature vector};

\node[smallbox, below left=1.2cm and 0.8cm of features] (cluster) {K-means Clustering\\
$k=3$ + PCA};
\node[smallbox, below right=1.2cm and 0.8cm of features] (ml) {Supervised ML\\
Logistic Regression,\\ Random Forest, MLP};

\draw[arrow] (corpus) -- (prep);
\draw[arrow] (prep) -- (length);
\draw[arrow] (prep) -- (freq);

\draw[arrow] (length) -- (ordinal);
\draw[arrow] (freq) -- (ordinal);

\draw[arrow] (ordinal) -- (static);
\draw[arrow] (ordinal) -- (dynamic);
\draw[arrow] (ordinal) -- (info);

\draw[arrow] (static) -- (features);
\draw[arrow] (dynamic) -- (features);
\draw[arrow] (info) -- (features);

\draw[arrow] (features) -- (cluster);
\draw[arrow] (features) -- (ml);

\end{tikzpicture}
}
\caption{Pipeline of the proposed methodology, from text preprocessing and ordinal pattern extraction to feature construction, K-means Clustering, and Supervised Machine Learning, Logistic Regression, Random Forest and Multilayer Perceptron}
\label{fig:pipeline}
\end{figure}

We use two numerical representations to analyze the texts through ordinal patterns and information-theoretic measures. One is based on word length and the other is based on lexical frequency. Both followed the same pipeline, from preprocessing and ordinal pattern extraction to feature construction, clustering, and supervised classification. Figure~\ref{fig:pipeline} summarizes the entire workflow.

The analysis began with a preprocessing stage aimed at obtaining clean and comparable textual sequences. For each document, all words were converted to lowercase and punctuation marks, numbers, accents, and special characters were removed. The texts were then tokenized at the word level. During this process, the original order of the tokens was preserved, since the subsequent ordinal analysis depends on the sequential organization of words in the text.

Stopwords were removed to reduce the influence of highly frequent terms with limited structural contribution. Since we want to preserve the sequential and structural properties of the original texts, we do not apply lemmatization or stemming.

Following the preprocessing stage, we constructed two complementary numerical representations. In the first, each word was replaced by its length, measured as the number of characters composing the word. This length-based representation provides a purely structural view of the text, independent of semantic content. In the second, each word was replaced by its relative frequency within the document. This frequency-based representation incorporates information about lexical repetition and word distribution. Both approaches were processed independently using the same analytical pipeline, allowing the comparison of two different but complementary ways of encoding textual structure.


Ordinal patterns were extracted from the numerical sequences following the method proposed by Bandt and Pompe \cite{bandt2002}, which transforms numerical subsequences into symbolic patterns according to the relative ordering of their values rather than their absolute magnitudes.

We use an embedding dimension of \(D = 3\) and a delay parameter of \(\tau = 1\). With this configuration, each subsequence of three values was mapped to one of \(3! = 6\) possible ordinal patterns:
\[
012,\ 021,\ 102,\ 120,\ 201,\ 210.
\]

For each document, the numerical sequence was divided into consecutive blocks of length three, and each block was transformed into its corresponding ordinal pattern. We extracted two types of information from the resulting symbolic sequences. The absolute count and probability of each pattern provided a global description of its distribution, and transitions between consecutive patterns were recorded to capture the local dynamics of the sequence.


To characterize the dynamics of the ordinal sequences, transition matrices were constructed for each document. Each matrix records how often a given ordinal pattern is followed by another. These transition counts were normalized to obtain transition probabilities, allowing the dynamics to be interpreted probabilistically.

In addition to transition-based features, we computed information-theoretic measures from the ordinal pattern probability distribution \(P=\{p_i\}_{i=1}^{D!}\) for each document \cite{bandt2002,rosso2007,LopezRuiz1995}. Permutation entropy was used to quantify the degree of uncertainty or randomness in the distribution of ordinal patterns and is defined as Bandt and Pompe \cite{bandt2002}.

\begin{equation}
H[P]=\frac{S[P]}{S_{\max}},
\end{equation}
where $S_{\max} = \ln(D!)$ is the Shannon entropy of the uniform distribution over the $D!$ possible ordinal patterns, so that $H[P] \in [0,1]$ and,

\begin{equation}
S[P]=-\sum_{i=1}^{D!} p_i \ln p_i.
\end{equation}

Disequilibrium measures the deviation of the observed distribution from the uniform distribution \(P_e\), indicating whether some ordinal patterns are more dominant than others. It is computed from the normalized Jensen-Shannon divergence \cite{rosso2007} as

\begin{equation}
Q[P]=Q_0\,J(P,P_e),
\end{equation}
where

\begin{equation}
J(P,P_e)=
S\!\left(\frac{P+P_e}{2}\right)
-\frac{S[P]}{2}
-\frac{S[P_e]}{2},
\end{equation}
and \(Q_0\) is the normalization constant that guarantees \(0\le Q[P]\le1\).

Finally, statistical complexity combines entropy and disequilibrium,
\begin{equation}
C[P]=H[P]\,Q[P],
\end{equation}
capturing the balance between regularity and variability in the ordinal structure of the text \cite{rosso2007}. These measures provide a compact description of the organization of each document and were used to analyze the relationship between entropy and complexity across the corpus.

For each document, we constructed a feature vector by combining the output of the ordinal analysis. The final representation included the total number of words, the absolute counts and probabilities of ordinal patterns, the transition probabilities between patterns, as well as permutation entropy, disequilibrium, and statistical complexity. The same procedure was applied to both numerical representations.




We analyzed the resulting feature vectors using K-means clustering. The number of clusters was determined by evaluating the average silhouette coefficient for values of $k$ ranging from 2 to 10 \cite{rousseeuw1987}. Based on this analysis, $k=3$ was selected for both numerical representations to ensure consistency and comparability between the length-based and frequency-based approaches. The silhouette analysis used for selecting $k$ is provided in the Supplementary Material (Figure~\ref{fig:k_selection}). Principal Component Analysis (PCA) was then applied to project the feature vectors into a two-dimensional space for visualization.



After the clustering stage, the cluster labels obtained from K-means were used as the target variable for supervised learning. The aim was to evaluate whether the ordinal-based feature vectors contained sufficient information to discriminate between the identified clusters. The dataset was divided into 80\% for training and 20\% for testing. The evaluated models included Logistic Regression, Random Forest, and a Multilayer Perceptron (MLP). A Dummy Classifier was also included as a baseline for comparison.

Model performance was evaluated on the test set using standard classification metrics, including accuracy, balanced accuracy, and macro F1-score. Additional experiments were conducted using reduced subsets of features to assess the contribution of different ordinal pattern descriptors and the robustness of the proposed representations.

\section{Results}

\begin{figure}[H]
    \centering

    \begin{subfigure}{0.48\textwidth}
        \centering
        \includegraphics[width=\linewidth]{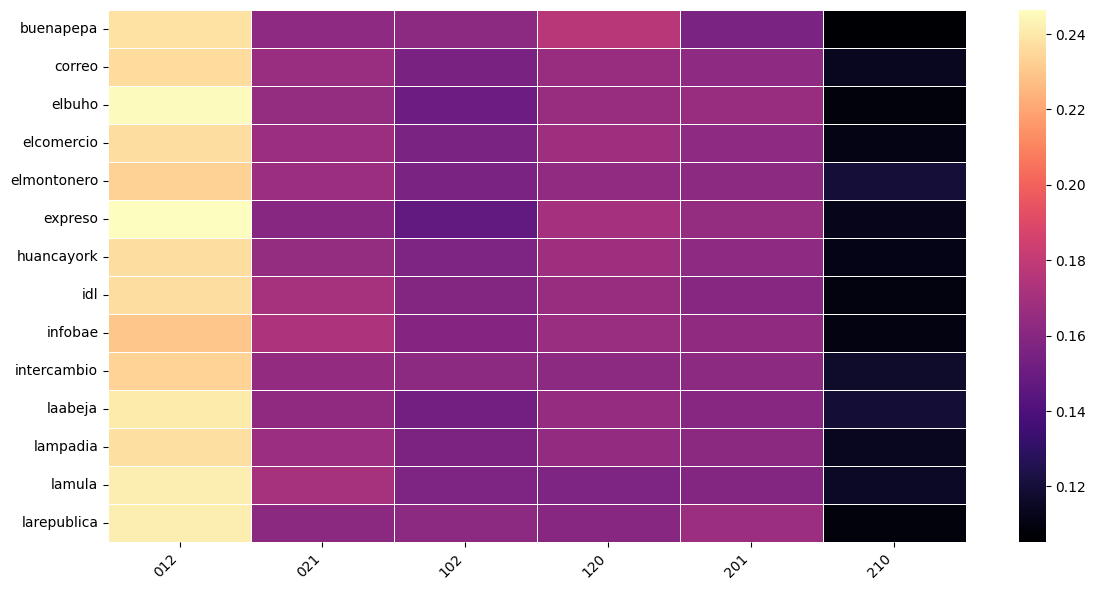}
        \caption{Length-based approach.}
        \label{fig:ordinal_patterns_e1}
    \end{subfigure}
    \hfill
    \begin{subfigure}{0.48\textwidth}
        \centering
        \includegraphics[width=\linewidth]{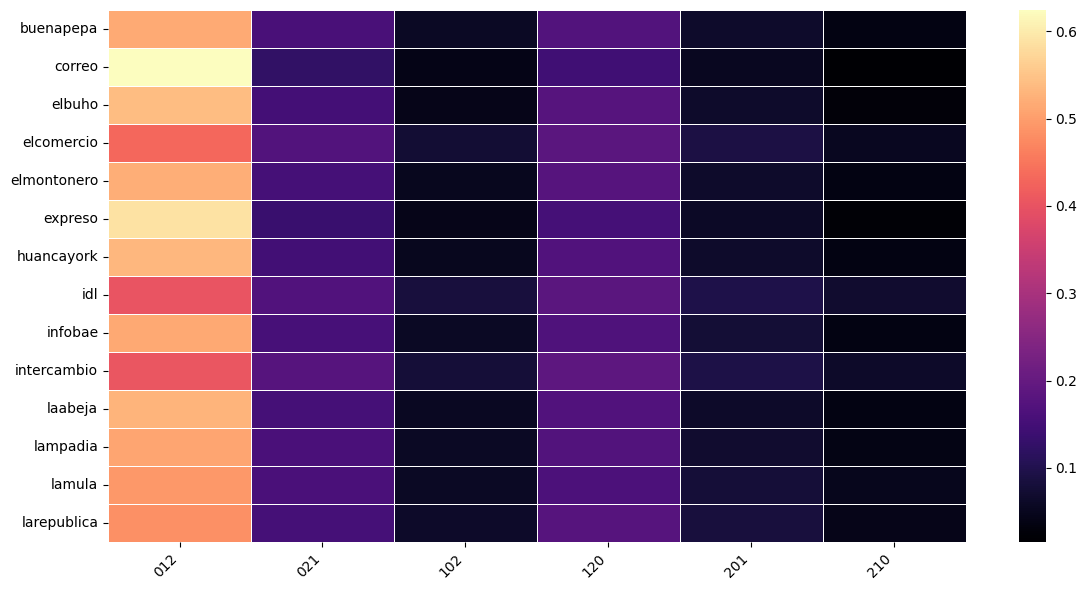}
        \caption{Frequency-based approach.}
        \label{fig:ordinal_patterns_e2}
    \end{subfigure}

    \caption{Heatmaps of ordinal pattern probabilities for the two numerical representations. Rows correspond to newspaper sources, columns correspond to ordinal patterns, and the color intensity indicates the probability of occurrence of each ordinal pattern. (a) Length-based representation. (b) Frequency-based representation.}
    \label{fig:ordinal_patterns}
\end{figure}

We show the ordinal pattern probability distributions for both representations in Fig.~\ref{fig:ordinal_patterns_e1} and Fig.~\ref{fig:ordinal_patterns_e2}. In these heatmaps, the x-axis corresponds to the ordinal patterns and the y-axis to the newspaper sources. The color intensity represents the probability of occurrence of each ordinal pattern, computed as the number of occurrences of a given pattern across all articles from a newspaper divided by the total number of ordinal patterns extracted for that newspaper.

The heatmaps in Fig.~\ref{fig:ordinal_patterns} reveal that ordinal patterns are not uniformly distributed under either numerical representation. In both cases, pattern $012$ is the most frequent and pattern $210$ is the least frequent across all newspaper sources,suggesting that both representations capture similar underlying structural properties despite their different numerical encodings. However, the two representations exhibit distinct distributional profiles. In the length-based representation (Fig.~\ref{fig:ordinal_patterns}a), the probabilities are relatively balanced across the six ordinal patterns, with $021$,$102$,$201$ showing comparable frequencies. In constrast, the frequency-based representation (Fig.~\ref{fig:ordinal_patterns}b) is more concentrated around a smaller subset of patterns. After the dominant pattern $012$, patterns $021$ and $120$ are the next most frequent, followed by $102$ and $201$, while pattern $210$ remains the least frequent.



\begin{figure}[H]
    \centering

    \begin{subfigure}{0.48\textwidth}
        \centering
        \includegraphics[width=\linewidth]{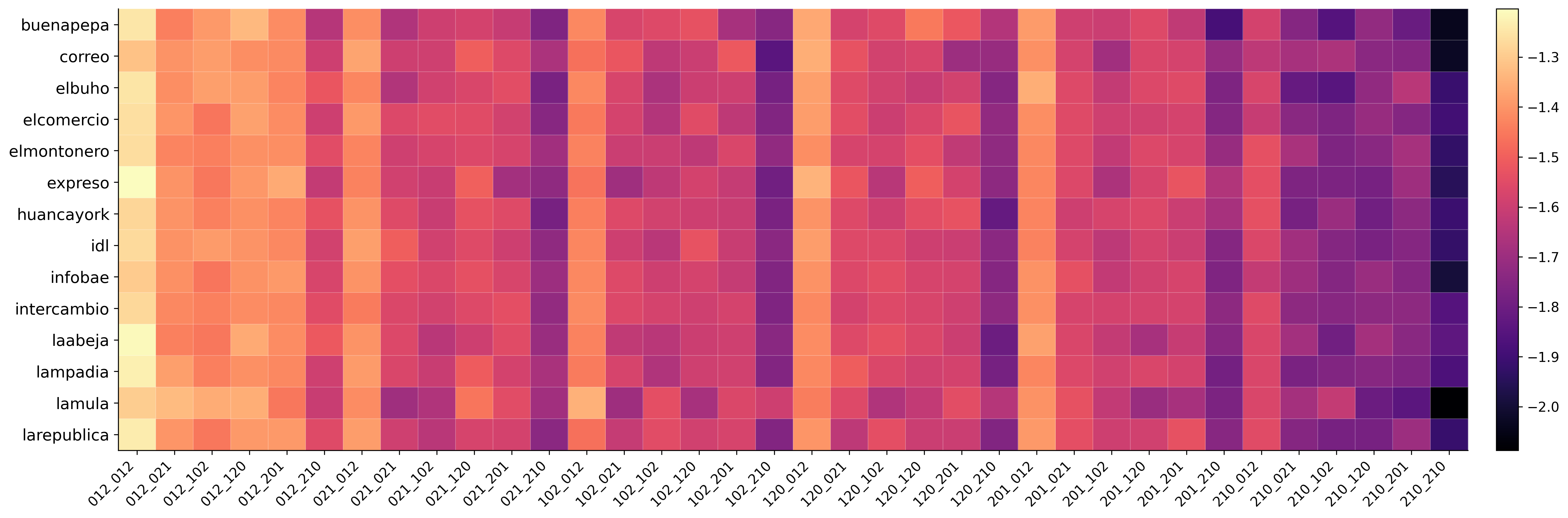}
        \caption{Length-based approach.}
        \label{fig:transition_matrix_log_e1}
    \end{subfigure}
    \hfill
    \begin{subfigure}{0.48\textwidth}
        \centering
        \includegraphics[width=\linewidth]{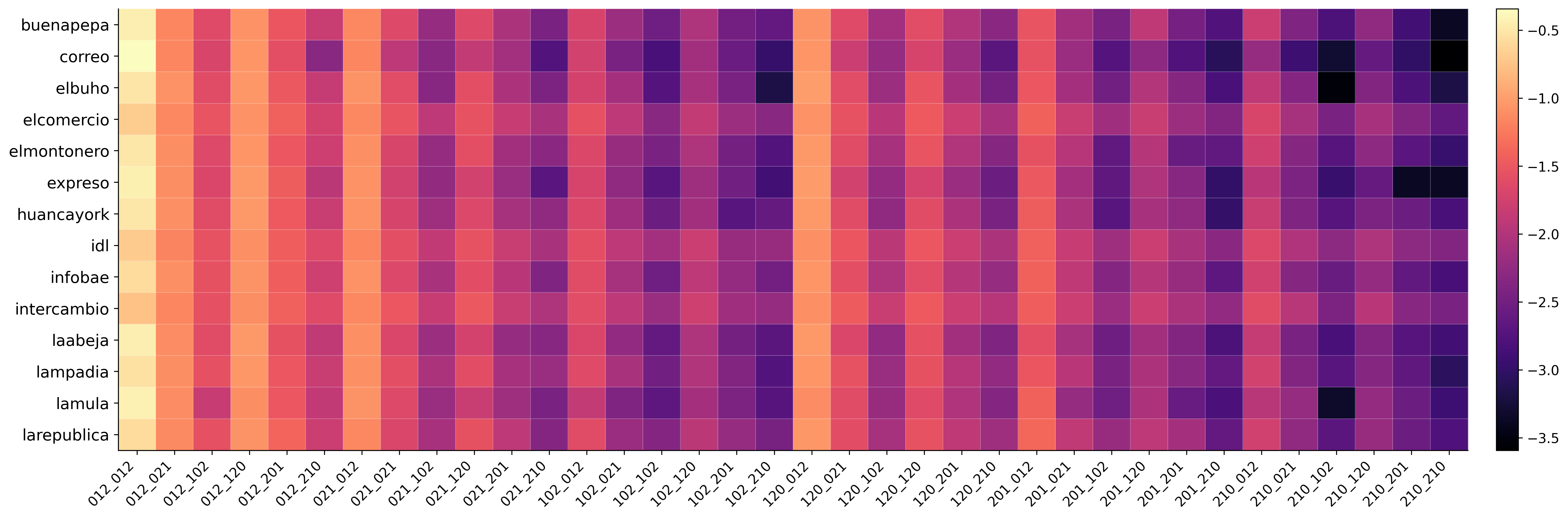}
        \caption{Frequency-based approach.}
        \label{fig:transition_matrix_log_e2}
    \end{subfigure}

    \caption{Log$_{10}$-transformed transition probability matrices for the two numerical representations. The x-axis represents the 36 possible transitions between consecutive ordinal patterns, and the y-axis corresponds to the analyzed newspapers. Cell colors indicate the log$_{10}$ of the transition probabilities, enhancing the visualization of low-probability transitions. (a) Length-based representation. (b) Frequency-based representation.}
    \label{fig:transition_matrix_log}
\end{figure}

We present the log$_{10}$-transformed transition probability matrices for both numerical representations in Fig.~\ref{fig:transition_matrix_log}. The x-axis displays the 36 possible transitions between consecutive ordinal patterns, while the y-axis corresponds to the analyzed newspapers. The color scale represents the log$_{10}$ of the transition probabilities, allowing low-probability transitions to be more clearly visualized. Since transition probabilities span several orders of magnitude, a log$_{10}$ transformation was applied to improve the visualization of the full probability range.

The transition probability matrices exhibit similar overall structures under both numerical representations. In both cases, the highest transition probabilities are associated with transitions involving pattern $012$, particularly the self-transition $012\!\rightarrow\!012$, whereas transitions involving pattern $210$, especially $210\!\rightarrow\!210$, have the lowest probabilities. Although the dominant transitions are similar in both representations, the frequency-based matrix (Fig.~\ref{fig:transition_matrix_log}b) exhibits stronger visual contrast between high- and low-probability transitions than the length-based matrix (Fig.~\ref{fig:transition_matrix_log}a).





The existence of these preferential trajectories, particularly the frequent recurrence of transitions involving pattern $012$ (e.g., $012\!\rightarrow\!012$), suggests that texts exhibit an internally structured dynamics when analyzed from an ordinal perspective. In both approaches, recurrent transition configurations can be identified, reinforcing the hypothesis that local textual organization contains non-trivial structural dependencies.



\begin{figure}[H]
    \centering
    \label{fig:entropy_complexity}
    \begin{subfigure}{0.48\textwidth}
        \centering
        \includegraphics[width=\linewidth]{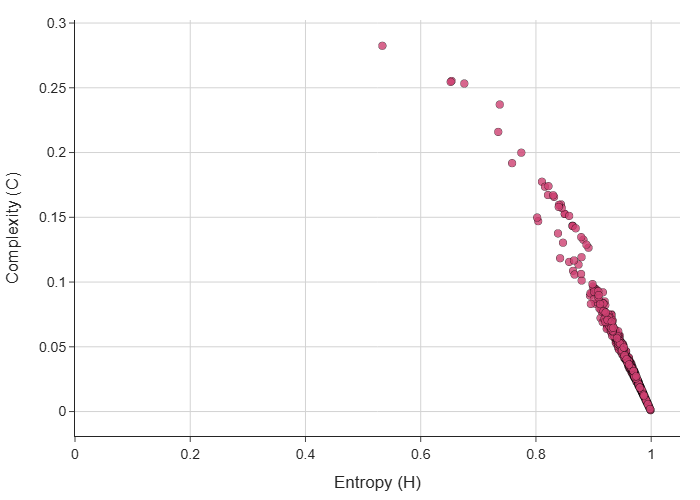}
        \caption{length-based approach}
        \label{fig:entropy_complexity_e1}
    \end{subfigure}
    \hfill
    \begin{subfigure}{0.48\textwidth}
        \centering
        \includegraphics[width=\linewidth]{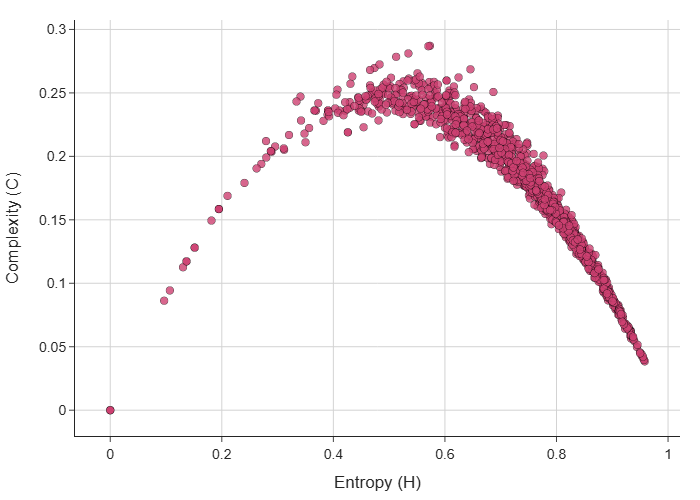}
        \caption{frequency-based approach}
        \label{fig:entropy_complexity_e2}
    \end{subfigure}

    \caption{Entropy-complexity planes obtained from the proposed representations. The x-axis represents the permutation entropy ($H$), while the y-axis represents the statistical complexity ($C$). Each point corresponds to one news article represented in the entropy-complexity space.}
    \label{fig:entropy_complexity}
\end{figure}

The entropy-complexity planes exhibit a characteristic relationship in which statistical complexity reaches its highest values at intermediate entropy levels. In Fig.~\ref{fig:entropy_complexity}, we illustrate the entropy versus complexity relationship for both
the length-based (Fig.~\ref{fig:entropy_complexity_e1}) and frequency-based (Fig.~\ref{fig:entropy_complexity_e2}) approaches.

This behavior is consistent with the entropy-complexity framework used in ordinal-pattern analysis, where the highest complexity values are typically associated with intermediate entropy levels, reflecting the coexistence of regularity and variability rather than complete order or complete randomness \cite{rosso2007,Zanin2012, herrera2020}.


The frequency-based representation (Fig.~\ref{fig:entropy_complexity_e2}) tends to concentrate at higher entropy values, whereas the length-based representation (Fig.~\ref{fig:entropy_complexity_e1}) shows a wider dispersion in statistical complexity values. This difference may be explained by the nature of the numerical encodings. Word lengths constitute a relatively stable structural property of the language, leading to similar ordinal distributions across documents. In contrast, lexical frequencies are more sensitive to vocabulary usage and repetition patterns, introducing greater variability in the ordinal sequences and consequently a broader range of entropy-complexity values. Despite these differences, the overall shape of the entropy-complexity relationship remains consistent across both representations, indicating that they capture complementary aspects of the underlying textual organization.

\subsection{Structure in the Feature Space}


We analyzed the resulting feature vectors using the K-means clustering algorithm as implemented in scikit-learn library \cite{pedregosa2011}. In Fig.~\ref{fig:clusters} we show the PCA projections obtained for both approaches. Texts can be grouped into well-defined regions, revealing the presence of latent structure in the representations derived from ordinal patterns.

\begin{figure}[H]
    \centering
    \label{fig:clusters}
    \begin{subfigure}{0.48\textwidth}
        \centering
        \includegraphics[width=\linewidth]{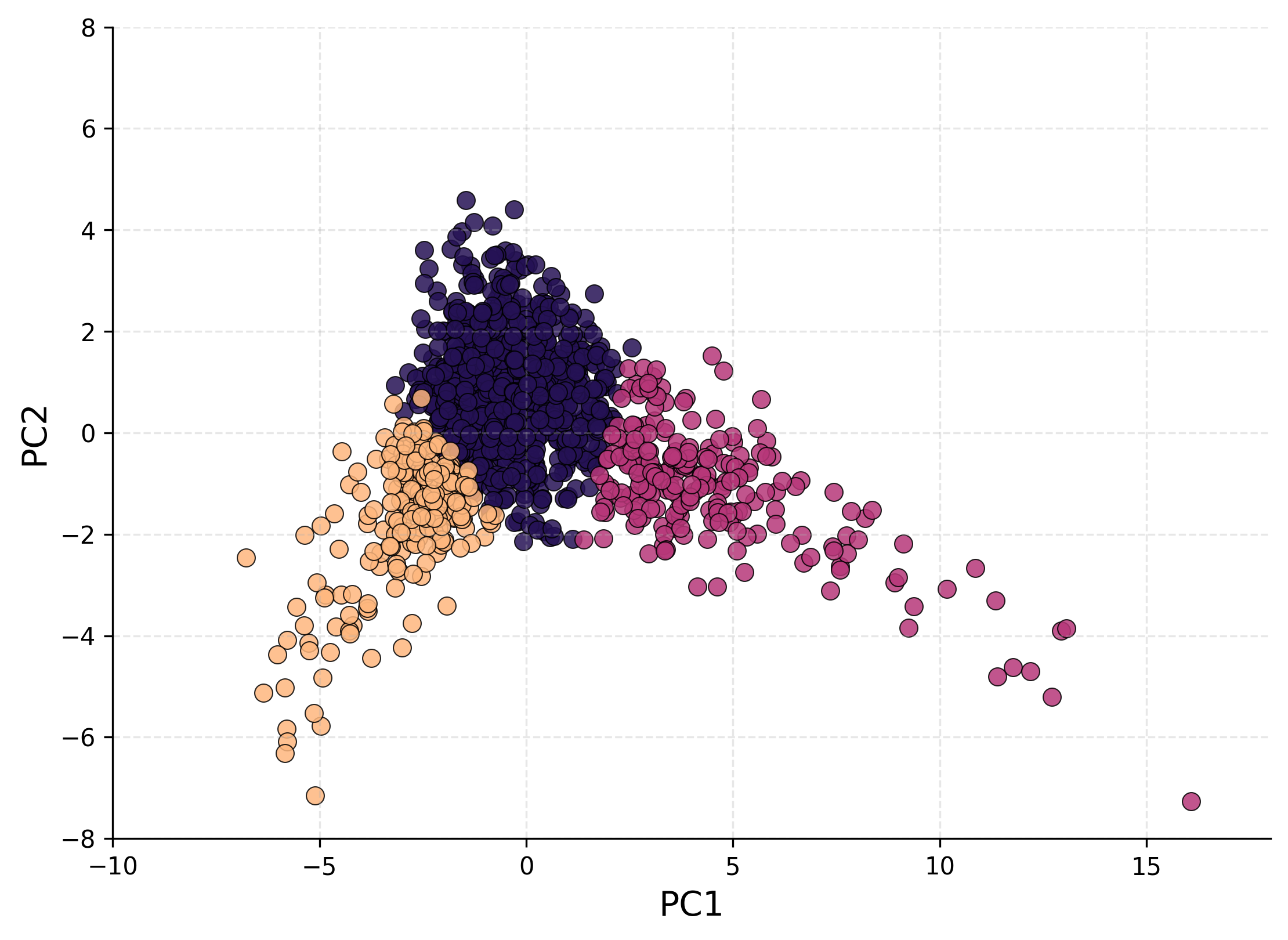}
        \caption{length-base approach}
        \label{fig:clusters_e1}
    \end{subfigure}
    \hfill
    \begin{subfigure}{0.48\textwidth}
        \centering
        \includegraphics[width=\linewidth]{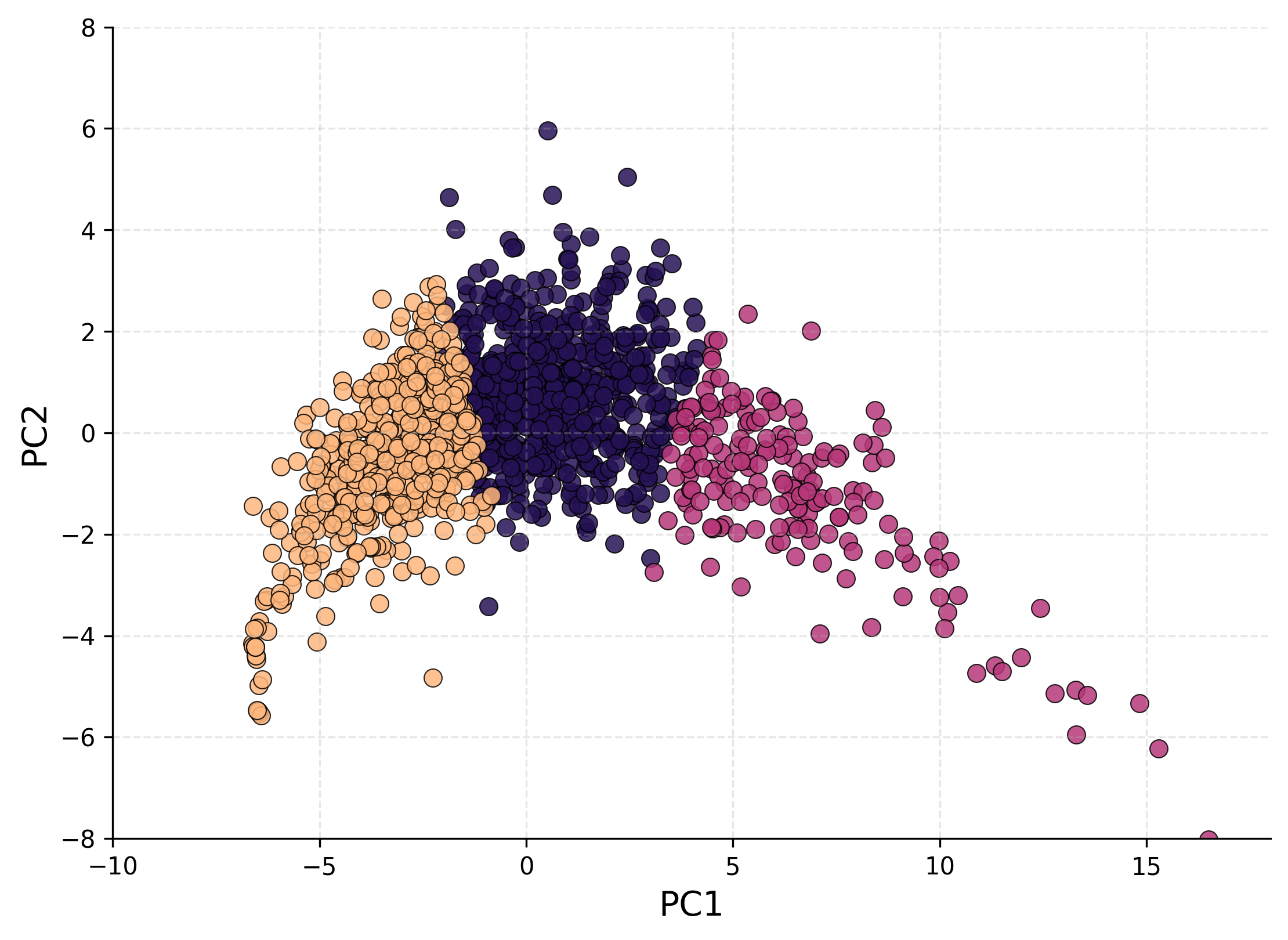}
        \caption{frequency-based approach}
        \label{fig:clusters_e2}
    \end{subfigure}

    \caption{PCA projection of the feature vectors extracted from ordinal pattern descriptors, with points colored according to the clusters identified by the K-Means algorithm. Each point represents one article in the reduced two-dimensional feature space defined by the first two principal components (PC1 and PC2).}
    \label{fig:clusters}
\end{figure}

In both approaches, K-means identified three clusters, indicating that the extracted features contain sufficient information to partition the texts into distinct groups. Although the separation is not strictly linear in the PCA projection, both numerical representations reveal comparable clustering patterns.

Table~\ref{tab:cluster_metrics} summarizes the clustering quality metrics for both representations. Compared with the length-based representation, the frequency-based representation achieves a higher silhouette score \cite{rousseeuw1987}, a lower Davies-Bouldin index \cite{davies1979}, and a higher Calinski-Harabasz score \cite{calinski1974}. These results suggest that lexical frequency provides a slightly more discriminative feature space while preserving the overall grouping structure.

\begin{table}[H]
\centering
\caption{Comparison of clustering quality metrics for the two numerical representations. Higher values indicate better clustering quality for the Silhouette and Calinski-Harabasz scores, whereas lower values indicate better quality for the Davies-Bouldin index. Better values for all metrics are in bold.}
\label{tab:cluster_metrics}
\begin{tabular}{lcc}
\toprule
\textbf{Metric} & \textbf{Length-based} & \textbf{Frequency-based} \\
\midrule
Silhouette score         & 0.0661 & \textbf{0.1115} \\
Davies-Bouldin index    & 2.9963 & \textbf{2.3438} \\
Calinski-Harabasz score & 104.77 & \textbf{203.07} \\
\bottomrule
\end{tabular}
\end{table}


The results obtained using supervised learning models confirm the discriminative capacity of the proposed representations \cite{savoy2020}. Models such as Logistic Regression and Random Forest achieve accuracy values close to or above 0.95 in both approaches.

\begin{table}[H]
\centering
\caption{Classification accuracy of the supervised learning models for the two numerical representations. Higher values indicate better accuracy and are highlighted in bold, except for the baseline dummy model.}
\label{tab:ml_results}
\renewcommand{\arraystretch}{1.15}
\setlength{\tabcolsep}{12pt}
\begin{tabular}{lcc}
\toprule
\textbf{Model} & \textbf{Length-based} & \textbf{Frequency-based} \\
\midrule
Dummy                    & 0.648 & 0.458 \\
Logistic Regression      & 0.984 & \textbf{0.992} \\
Random Forest            & 0.964 & \textbf{0.968} \\
RF without Top Feature   & 0.952 & \textbf{0.976} \\
RF with Top 5 Features   & \textbf{0.956} & \textbf{0.956} \\
MLP                      & 0.948 & \textbf{0.952} \\
\bottomrule
\end{tabular}
\end{table}

The two numerical representations achieve consistently high classification performance. Although the frequency-based representation attains slightly higher accuracy for most models, the differences are small, indicating that both representations provide similarly informative feature spaces for supervised classification.

Furthermore, even when the number of features is reduced, for instance, retaining only the most relevant variables, the performance remains consistently high. This suggests that discriminative information is not concentrated in a single feature, but distributed across the overall representation.

In contrast, the baseline dummy model exhibits significantly lower performance, confirming that ordinal patterns capture relevant and non-trivial information for text classification. Table~\ref{tab:ml_results} summarizes the classification performance obtained using supervised learning models in terms of accuracy. For Random Forest, we also obtained a Balanced Accuracy of 0.959 and Macro F1 of 0.957 for the length-base approach and a Balanced Accuracy of 0.976 and Macro F1 of 0.969 for the frequency-based approach.






\section{Discussion}

The analysis of Peruvian online newspapers shows that journalistic sources leave detectable structural fingerprints at the ordinal level \cite{amancio2015}, independently of semantic content. This is consistent with evidence from other domains where ordinal-pattern-based entropy and complexity measures have been shown to encode style-specific signatures without requiring semantic interpretation, such as in the classification of pictorial styles across art history \cite{sigaki2018}. The classification accuracy obtained across all models (Table~\ref{tab:ml_results}) directly supports the hypothesis that structural sequence properties alone are sufficient for reliable source attribution.

The dominance of pattern $012$ across all 14 newspaper sources in both encodings (Figs.~\ref{fig:ordinal_patterns_e1} and~\ref{fig:ordinal_patterns_e2}) reflects the syntactic structure of Spanish \cite{sanchez2023}. In the length-based encoding (Fig.~\ref{fig:ordinal_patterns_e1}), a total of 144,943 ordinal patterns were extracted from the corpus, of which 23.70\% correspond to pattern $012$. The remaining patterns occur with relatively similar frequencies, except for pattern $210$, which accounts for only 11.26\% of the extracted patterns. In contrast, the frequency-based encoding (Fig.~\ref{fig:ordinal_patterns_e2}) also contains 144,943 extracted ordinal patterns, but pattern $012$ represents 48.12\% of all patterns, whereas pattern $210$ accounts for only 4.46\%. The transition matrices (Figs.~\ref{fig:transition_matrix_log_e1} and~\ref{fig:transition_matrix_log_e2}) show that transitions involving pattern $012$ are consistently more probable than other transitions across all newspaper sources, whereas transitions involving pattern $210$ are the least frequent.


The entropy-complexity plane (Figs.~\ref{fig:entropy_complexity_e1} and~\ref{fig:entropy_complexity_e2}) reveals a meaningful difference between the two encodings. Under the word-length encoding, documents concentrate at high entropy values ($H > 0.8$) with low complexity (Fig.~\ref{fig:entropy_complexity_e1}), placing them near the random end of the plane. Under the lexical frequency encoding, documents trace the theoretical arc of complex systems \cite{rosso2007,stanisz2024}, a curve built from the interplay between entropy and disequilibrium \cite{LopezRuiz1995}, across a wide range of entropy values (Fig.~\ref{fig:entropy_complexity_e2}). Word-length sequences in journalistic text are therefore nearly random in their ordinal ordering, whereas word-frequency sequences exhibit genuine structural complexity. The two encodings do not contradict each other; rather, they describe complementary structural layers of the same texts.

The PCA projections (Figs.~\ref{fig:clusters_e1} and~\ref{fig:clusters_e2}) show three coherent groups in both encodings. This agrees with Pessa et al.~\cite{pessa2022}, who showed that unsupervised clustering over entropy-complexity representations can recover coherent behavioral groups without prior labeling, even in physical systems unrelated to language. One cluster displays substantially greater spread along the first principal component in both figures, suggesting higher internal variability among the corresponding sources. The partial overlap between the remaining two clusters indicates that some sources share ordinal properties regardless of their editorial content.

All supervised models substantially outperform the Dummy Classifier, which achieved accuracies of 0.648 and 0.458 for the length-based and frequency-based representations, respectively (Table~\ref{tab:ml_results}). Logistic Regression achieves the highest accuracy in both encodings (0.984 and 0.992), indicating that the feature space is largely linearly separable. Random Forest reaches accuracies of 0.964 and 0.968, balanced accuracies of 0.959 and 0.976, and macro F1-scores of 0.957 and 0.969. The MLP obtains accuracies of 0.948 and 0.952. Removing the single most influential variable reduces accuracy only to 0.952 and 0.976, whereas retaining only the top five features yields 0.956 in both encodings, confirming that discriminative information is distributed across the overall representation rather than concentrated in a single descriptor.

One of the central findings of this study is the consistency of the results obtained from the two numerical representations. Despite encoding texts through different structural properties—word length and lexical frequency—both approaches converge on the same overall conclusions. In both cases, ordinal pattern distributions are non-uniform, transition structures exhibit preferential trajectories, the entropy-complexity relationships reveal non-trivial organization, K-means clustering identifies coherent groups in the feature space, and supervised learning models achieve high classification performance.

Taken together, these findings suggest that the ability to discriminate between newspaper sources does not depend on a particular numerical representation, but rather on deeper structural properties of language \cite{ferrericancho2003,stanisz2024}. The consistency observed across both representations supports the hypothesis that ordinal-pattern-based features capture an intrinsic structural fingerprint of the texts rather than artifacts introduced by a specific encoding. Consequently, the length-based and frequency-based representations should be interpreted as complementary descriptions of the same underlying textual dynamics rather than competing alternatives.

The study is limited to a single linguistic and geographic context. Generalization to other languages, journalistic traditions, or media formats remains to be demonstrated. The choice of $D=3$ limits the ordinal vocabulary to six patterns; higher values of $D$ could capture longer-range structural dependencies, at the cost of requiring substantially longer texts for reliable estimation. Future work should identify which ordinal patterns and transitions contribute most to distinguishing newspaper sources, investigate how these structural properties vary across different article genres and text collections, and extend the proposed methodology to different types of texts, larger corpora, and other languages.

\section{Conclusions}

This study shows that ordinal pattern analysis and information-theoretic measures can effectively characterize and attribute Spanish-language journalistic texts at the structural level. This is particularly relevant since computational text analysis methods, including stylometric and ordinal-pattern approaches, have been disproportionately developed and validated on English-language corpora, leaving major world languages such as Spanish comparatively underexplored \cite{joshi2020}. Applied to news articles from Peruvian online newspapers, both the word-length and lexical frequency encodings yield non-uniform ordinal distributions, structured transition matrices, and distinct entropy-complexity signatures across sources. Supervised classification models trained on these features achieve accuracy above 0.95 in all configurations, with performance remaining stable even under significant feature reduction. The consistency of these results across two structurally different encodings supports the view that ordinal features capture a genuine property of each source's writing style, not an artifact of any particular representation.

These findings extend the application of ordinal pattern methods to a non-English journalistic setting and confirm that source attribution can be carried out from structural sequence properties alone. Beyond its theoretical interest, structural source attribution has practical relevance in contexts such as verifying content provenance and distinguishing human-written from machine-generated news, a growing concern as large language models are increasingly used to produce journalistic-style text \cite{iaquinta2024}. Future work should examine whether these fingerprints hold across other languages, media formats, and time periods, and investigate the linguistic interpretation of the most discriminative ordinal transitions, as suggested by S\'anchez \cite{sanchez2023}, linking ordinal pattern distributions to typological differences across languages. 
\bibliographystyle{ieeetr}
\bibliography{referencias}

\newpage
\appendix

\setcounter{table}{0}
\setcounter{figure}{0}

\section{Peruvian online newspapers}
\label{sec:apendiceA}

We scrapped fourteen different Peruvian online newspapers between May and June 2025, as shown in Table~\ref{tab:S1}. Their choice was based on availability, but also because they are the most known newspapers in Perú with opinion columns.

\begin{table}[ht]
\centering
\caption{Peruvian online newspapers included in the corpus and their approximate access dates.}
\label{tab:S1}
\begin{tabular}{ll}
\toprule
\textbf{Newspaper} & \textbf{Accessed} \\
\midrule
Buenapepa      & \multirow{14}{*}{May-June 2025} \\
Correo         & \\
El Búho        & \\
El Comercio    & \\
El Montonero   & \\
Expreso        & \\
Huancayork     & \\
IDL Reporteros & \\
Infobae Perú   & \\
Intercambio    & \\
La Abeja       & \\
Lampadia       & \\
La Mula        & \\
La República   & \\
\bottomrule
\end{tabular}
\end{table}


\section{Additional PCA projections}



\begin{figure}[H]
\centering

\begin{subfigure}{0.48\textwidth}
    \centering
    \includegraphics[width=\linewidth]{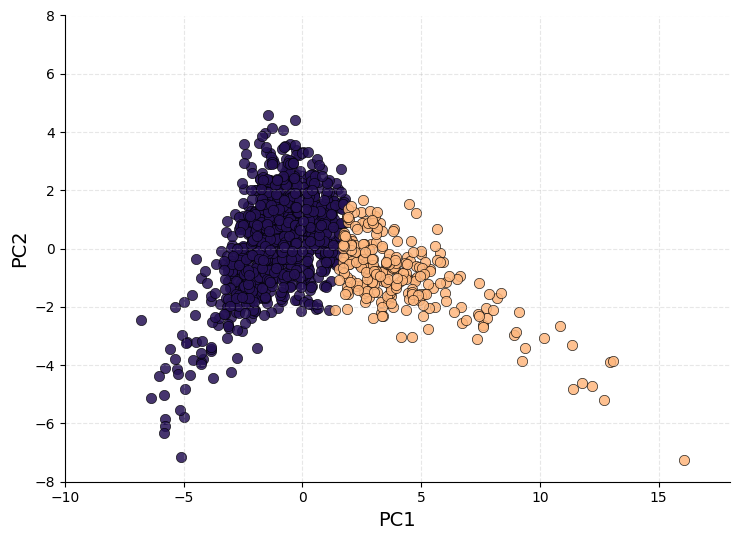}
    \caption{Length-based approach (\(k=2\)).}
\end{subfigure}
\hfill
\begin{subfigure}{0.48\textwidth}
    \centering
    \includegraphics[width=\linewidth]{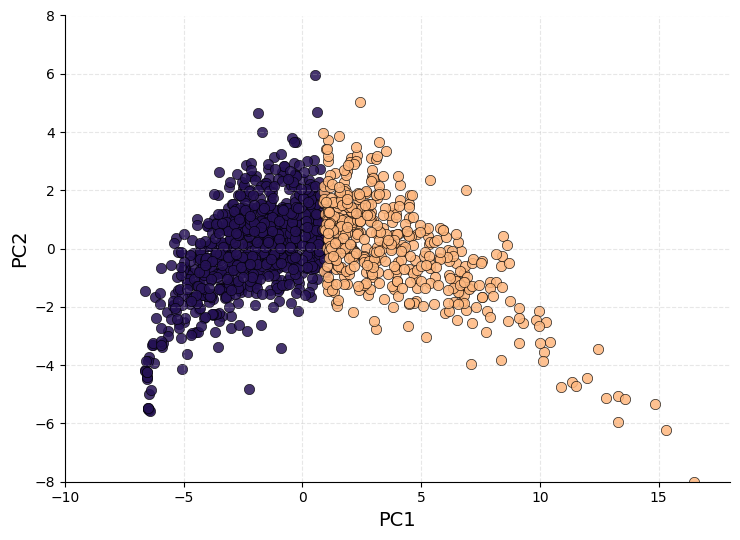}
    \caption{Frequency-based approach (\(k=2\)).}
\end{subfigure}

\vspace{0.4cm}

\begin{subfigure}{0.48\textwidth}
    \centering
    \includegraphics[width=\linewidth]{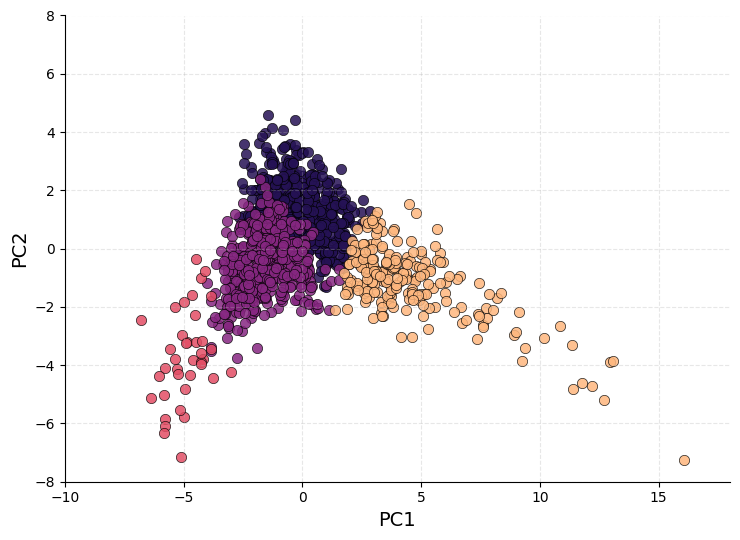}
    \caption{Length-based approach (\(k=4\)).}
\end{subfigure}
\hfill
\begin{subfigure}{0.48\textwidth}
    \centering
    \includegraphics[width=\linewidth]{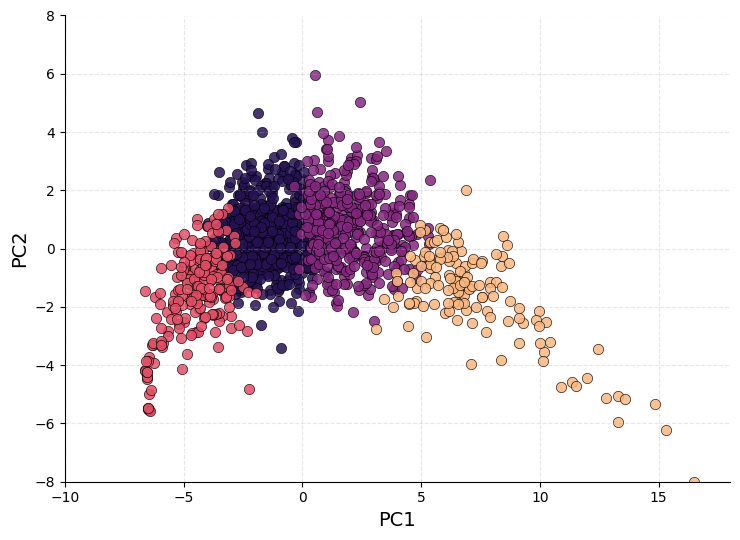}
    \caption{Frequency-based approach (\(k=4\)).}
\end{subfigure}

\caption{PCA projections of the feature vectors obtained using K-means clustering with \(k=2\) and \(k=4\) for the two numerical representations. Each point represents one news article projected onto the first two principal components (PC1 and PC2). For both representations, the \(k=2\) solution partitions the corpus into two broad groups, whereas the \(k=4\) solution produces a finer partition of the feature space by splitting existing groups into smaller clusters. These comparisons support the selection of \(k=3\) for the analyses presented in the main text.}

\label{fig:k_selection}
\end{figure}

Figure~\ref{fig:k_selection} compares the PCA projections obtained with \(k=2\) and \(k=4\) for the length-based and frequency-based representations. For both approaches, the \(k=2\) solution partitions the corpus into only tow broad groups. whereas \(k=4\) solution further subdivides the feature space into smaller clusters. Together, these comparisons illustrate that \(k=3\) provides a balanced partition of the data, avoiding both under partitioning (\(k=2\)) and over partitioning (\(k=4\)) while preserving comparable clustering behavior across the two numerical representations.

\end{document}